\documentclass[aps,twocolumn,showpacs]{revtex4}

\usepackage{graphicx,amsmath,color}

\newcommand{\eq}[1]{Eq.~(\ref{#1})}
\newcommand{\fig}[1]{Fig.~\ref{#1}}

\newcommand{\olcite}[1]{Ref.~\onlinecite{#1}}
\newcommand{\olcites}[1]{Refs.~\onlinecite{#1}}

\newcommand{\eeq}{ \end{equation} }
\newcommand{\beq}{ \begin{equation} }

\newcommand{\eea}{ \end{eqnarray} }
\newcommand{\bea}{ \begin{eqnarray} }

\newcommand{\etal}{ {\em et al.}}
\newcommand{\bhua}{ {\bf \hat{u}}_{1} }
\newcommand{\bhub}{ {\bf \hat{u}}_{2} }

\newcommand{\bhu}{ {\bf \hat{u}} }
\newcommand{\aza}{  \psi_{1} }
\newcommand{\azb}{  \psi_{2} }
\newcommand{\ta}{  t_{1} }
\newcommand{\tb}{  t_{2} }
\newcommand{\qa}{  k_{1} }
\newcommand{\qb}{  k_{2} }

\newcommand{\lang} {\left \langle \left \langle }
\newcommand{\rang} {\right \rangle \right \rangle}

\newcommand{\bra}{ {\bf r}_1 }
\newcommand{\brb}{ {\bf r}_2 }
\newcommand{\br}{ {\bf r} }
\newcommand{\bz}{ {\bf \hat{z}} }
\newcommand{\bn}{ {\bf \hat{n}} }
\newcommand{\bx}{ {\bf \hat{x}} }
\newcommand{\by}{ {\bf \hat{y}} }
\newcommand{\bv}{ {\bf \hat{v}} }
\newcommand{\bw}{ {\bf \hat{w}} }

\begin{document}

\title{Cholesteric order in systems of helical Yukawa rods }
\author{H. H. Wensink}
\altaffiliation{Present address: Institute for Theoretical Physics:
  Soft Matter, Heinrich-Heine-University-D\"{u}sseldorf,
  Universit\"{a}tsstrasse 1, 40225, D\"{u}sseldorf, Germany}

\email{wensink@thphy.uni-duesseldorf.de}
\author{G. Jackson}
\affiliation{Department of Chemical Engineering, Imperial College London, South Kensington Campus, London SW7 2AZ, United Kingdom}

\date{\today}

\begin{abstract}

We consider the interaction potential between two chiral rod-like
colloids which consist of a thin cylindrical backbone decorated  with a
helical charge distribution on the cylinder surface. For sufficiently
slender coiled rods a simple scaling expression
is derived which links the
chiral `twisting' potential to the intrinsic properties of the particles such as the
coil pitch, charge density and electrostatic
screening parameter.  To predict the behavior of the macroscopic cholesteric pitch  
we invoke a simple second-virial theory generalized for weakly twisted
director fields. While the handedness of the cholesteric phase
for weakly coiled rods is always commensurate  with that of the internal
coil,  more strongly coiled rods display cholesteric order with
opposite handedness. The correlation
between the symmetry of the microscopic helix and the macroscopic
cholesteric director field is quantified in detail.  Mixing helices with sufficiently
disparate lengths and coil pitches gives rise to a demixing 
of the uniform cholesteric phase into two fractions with
a different macroscopic pitch. Our findings are consistent with
experimental results and could be helpful in interpreting experimental observations in systems of cellulose and
chitin microfibers, DNA and {\em fd} virus rods.

\end{abstract}

\pacs{61.30.Cz, 64.70.M, 82.70.Dd}

\maketitle

\section{Introduction}

In contrast to a common nematic phase, where the nematic
director is homogeneous throughout the system, the cholesteric (chiral nematic) phase is
characterized by a helical arrangement of the director field along a
common pitch axis. As a result, the cholesteric phase possesses an
additional mesoscopic length scale, commonly referred to as the
`cholesteric pitch ', which
characterizes the distance along the pitch axis over which the local director
makes a full turn \cite{gennes-prost}.
The behavior of the pitch as a function of density, temperature
and solvent conditions is of great fundamental and practical
importance since the unique rheological, 
electrical and optical properties of cholesteric materials are
determined in large part by the topology of the nematic director field.

In recent years considerable research effort has been devoted to
studying chirality in lyotropic liquid crystals which consist of
colloidal particles or stiff polymers immersed in a solvent. 
In addition to a number of synthetic  helical polymers such as polyisocyanates \cite{aharoni,sato-sato} and polysilanes \cite{watanabe-kamee} which form cholesteric phases in organic solvents there is a large class of helical bio-polymers which are known to form cholesteric phases in water. Examples are  DNA
\cite{robinson-pblg,livolantDNAoverview} and the rod-like $fd$-virus
\cite{dogic-fraden_fil}, polypeptides \cite{uematsu,dupresamulski}, chiral micelles \cite{hiltrop},
polysaccharides \cite{sato-teramoto}, and microfibrillar cellulose (and its derivatives)
\cite{gray-cellulose} and chitin \cite{chitin-revol}. In these systems, the
cholesteric pitch is strongly dependent upon the particle concentration, temperature as
well as e.g. the ionic strength which has been the subject of intense experimental research \cite{rill-dna,yu-dna,strey-dna, dogic-fraden_chol,grelet-fraden_chol,robinson-pblg, dupreduke75,yoshiba-sato,dong-gray-cellulose,miller-cellulose,microcellulose}.

Understanding the connection between the molecular interactions responsible for
chirality on the microscopic scale and the structure of the macroscopic cholesteric phase has been a long-standing challenge in the physics
of liquid crystals \cite{gennes-prost}. 
The chiral nature of most biomacromolecules originates from a  spatially
non-uniform distribution of charges and dipole moments residing on the
molecule. The most prominent example is the double-helix backbone structure of the phosphate
groups in DNA. 
Combining the electrostatic interactions with the intrinsic
conformation of the molecule  allows for a coarse-grained
description in terms of an {\em effective} chiral shape. 
Examples are bent-core or banana-shaped
molecules \cite{straley,jakli} where the mesogen shape is primarily responsible for
chirality. Many other helical bio-polymers and microfibrillar
assemblies of chiral molecules (such as cellulose) can be mapped onto effective
chiral objects such as a threaded cylinder \cite{straley,odijkchiral,kimura2},
twisted rod \cite{chitin-revol,orts-cellulose} or semi-flexible helix \cite{pelcovits}.

The construction of a  microscopic theory for the cholesteric phase is
a serious challenge owing to the complexity of the underlying chiral interaction and the inhomogeneous and anisotropic nature of the phase \cite{harris-rmp}.
Course-grained model potentials aimed at capturing the essentials of the complex molecular nature of the electrostatics of the surface of such macromolecules have been devised mainly for DNA  \cite{parsegian-podgornik,kornyshevleikin-chiral-err,kornyshevleikin-prl, kornyshevleikin-pitch,tombolatoferrarini}. A more general electrostatic model potential for chiral interactions was proposed much earlier by Goossens \cite{goossens} based on
a spatial arrangement of dipole-dipole and dipole-quadrupole interactions
which can be cast into a multipole expansion in terms of tractable pseudo-scalar
potentials \cite{meervertogen}. This type of electrostatic description
of the chiral interaction can be combined with a Maier-Saupe
mean-field treatment \cite{meervertogenJCP,lin-liu77,hu,osipov,kapanowski,emelyanenko}, or with a bare
hard-core model and treated within the seminal theory of Onsager \cite{onsager,vargachiral1,wensinkjackson}. 

A drawback of the Goossens potential is that its simple
scaling form precludes an explicit connection with the underlying microscopic (electrostatic)
interactions involved. As a
consequence, the potential is unsuitable for studying the
sensitivity of the cholesteric pitch of charged rod-like colloids
where the strength of the chiral interactions depends strongly on the
configuration of the surface charges and the
electrostatic screening ({\em viz.} salt concentration).
In this paper, we propose a simple but explicit model for long-ranged chiral
interactions based on an impenetrable cylindrical rod decorated with a helical
coil of charges located on the rod surface.
The local interactions between the charged helical segments is
represented by a simple Yukawa potential in line with the
Debye-H\"{u}ckel approximation valid for weakly charged poly-electrolytes.
A subsequent analysis of the chiral potential between a pair of helical Yukawa
segment rods leads to a simple general expression which relates the
intrinsic twisting potential  to the electrostatic screening, charge density and coil pitch
of the helices.  The overall form of the chiral Yukawa potential resembles the one proposed by
Goossens \cite{goossens}, albeit with a much weaker decay with respect
to the interrod distance. The most important difference, however, is that the amplitude of the chiral Yukawa potential can be linked 
to the coil configuration, and the local electrostatic potential. By mapping the helical
charge distribution onto an effective helical coil the present model
could be interpreted as a simple prototype for a charged twisted rod,
a model commonly invoked to explain cholesteric organization of
cellulose \cite{orts-cellulose,yi-grafted-cellulose} and chitin 
microfibers \cite{chitin-revol}. Likewise, the 
$\alpha$-helical structure of polypeptide molecules \cite{kimura2} could also be
conceived as a coiled rod on a coarse-grained level.

In the second part of the analysis, the potential will be employed in a simple Onsager second
virial theory to predict the behavior of the macroscopic
cholesteric pitch of Yukawa helices. The sensitivity of the
 pitch with respect to the coil configuration and amplitude of the electrostatic interactions will be scrutinized
 in detail. Also considered are the implications of mixing two helices species with
different lengths and/or coil pitches on the isotropic-cholesteric
phase behavior of mixed systems. 

The theoretical results are consistent with experimental findings and
correctly capture the response of the cholesteric pitch upon
variation of the particle concentration and salt content. The theory also
unveils a subtle relationship between the internal
conformation (coil pitch) of the helical rod and the handedness of the
cholesteric phase. The sensitive relationship between the symmetry
of the cholesteric phase and the charge pattern on the rod surface is consistent 
with  experimental  observations in filamentous virus systems such as {\em
  fd} \cite{barrybeller} and {\em M13} \cite{tombolato-grelet} as well as numerical calculations based on a more explicit
poly-electrolyte site model \cite{tombolato-grelet}. 

This paper is structured as follows. In  Section II we introduce
the chiral Yukawa model and derive a general expression for the
chiral potential imparted by the helical charge distribution. The implications of the potential for the macrosopic cholesteric pitch will be analyzed
in detail in Section III where we shall focus first on monodisperse systems of identical
helices followed by a treatment of binary mixtures of Yukawa helices  differing
in length and/or coil pitch. Finally, Section IV will be devoted to a
discussion followed by some concluding remarks.

\begin{figure}
\begin{center}
\includegraphics[clip=,width= 0.12\columnwidth ]{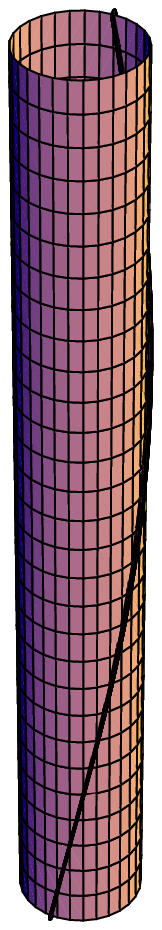}
\includegraphics[clip=,width= 0.12\columnwidth ]{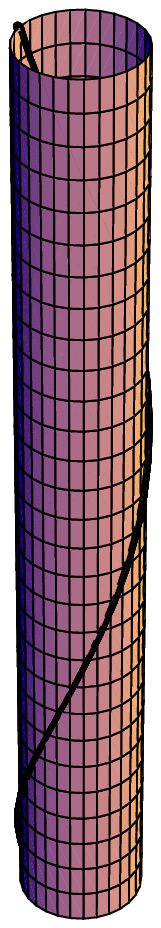}
\includegraphics[clip=,width= 0.12\columnwidth ]{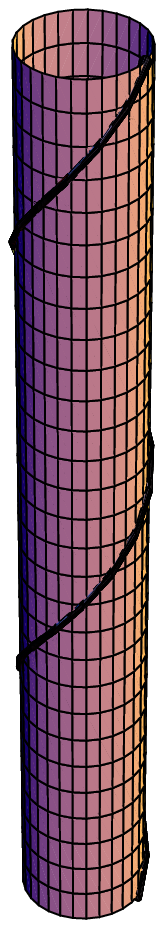}
\begin{picture}(0,0)
\put(1,1) {{\huge $\uparrow$}}
\put(1.1,1.8) {$ \bhu $}
\end{picture}
\caption{ \label{snap} Cylinders decorated with a continuous helical
  charge distribution (indicated by the thread) with different coil
  pitches, $k= \pi$ (left), $k = 2\pi$ (middle) and $k = 4 \pi$ (right).}
\end{center}
\end{figure}

\section{Chiral interaction between Yukawa helices}

In this study we aim to quantify the chiral or twisting potential of a pair of charged helical colloidal rods starting from a continuum electrostatic model based on a screened Coulombic (or Yukawa) potential. 
Within linearized Poisson-Boltzmann theory the electrostatic interaction between two charged point particles with equal charge $\pm Ze$ in a dielectric solvent with relative permittivity $\varepsilon_{r}$ is given by \cite{hansenmcdonald}:
\beq
\beta U_{Y}{(r)} = Z^{2}\lambda_{B} \frac{ \exp[- \kappa r]}{r} \label{yuk}
\eeq
with $r$ the distance between the particles, 
 $\lambda_{B} = e^{2}/4\pi \varepsilon_{0}\varepsilon_{r} k_{B}T$ the Bjerrum length ($\lambda_{B} = 0.7$ $nm$ for water at $T=298K$), $\varepsilon_{0}$ the vacuum permittivity, and $\kappa$ the Debye screening constant $\kappa = \sqrt{4 \pi \lambda_{B} (Z \rho + 2 c_{s})}$
which measures the extent of the {\em electric double layer} surrounding each particle. Here, $\rho = N/V$ is the 
macro-ion density, $N_{av}$ is Avogadro's number, and $c_{s}$ refers to the concentration of added salt. 

\eq{yuk} can be generalized in a straightforward manner to describe the interaction between two helical macro-ions by 
assuming that the total pair potential can be written as a summation
over Yukawa sites residing on the helices (see \fig{snap}). In the limit of an infinite 
number of sites per unit length the electrostatic potential of helical rods with variable lengths $L_{1}$ and $L_{2}$, and common diameter $D$ at positions $\{ \bra$ 
,$\brb \}$ and orientational unit vectors $\{ \bhua$ $\bhub \}$ is then given by a double contour integral over the longitudinal (pitch) axes of the helices:
\bea
\beta U_{Y} ( \Delta \br ; \bhua , \bhub ; \aza , \azb )  &=&
Z_{1}Z_{2} \lambda_{B}  \int_{-1}^{1} d \ta \int _{-1}^{1}  d \tb \nonumber \\ 
&& \times  \frac{ \exp[- \kappa |\br_{2}^{s} - \br_{1}^{s}| ]}{ |\br_{2}^{s} - \br_{1}^{s}|} \label{yukhel}
\eea
with  $\Delta \br = \brb - \bra$ the distance vector between the centre-of-masses of the helices, and $Z_{i}$ is the total number of charges on species $i$. 

In order to describe the segment position $\br_{i}^{s}$ along the contour of helix $i$ it is expedient to introduce a  particle-based Cartesian frame spanned by tree orthogonal unit vectors $\{ \bhu_{i} , \bv , \bw_{i} \}$ $(i=1,2)$, where
$\bv = \bhua \times \bhub  / |\bhua \times \bhub |$ and $\bw_{i} =
\bhu_{i} \times \bv $.  Within the particle the segment positions can be parametrized
as follows:
\begin{eqnarray}
  \br_{1}^{s} &=& \bra + \frac{L_{1}}{2} \ta  \bhua +  \frac{D}{2}  \nonumber \\
&& \times  \left \{  \cos ( \qa \ta  + \aza ) \bv + \sin ( \qa \ta + \aza ) \bw_{1} \right \} \nonumber \\
  \br_{2}^{s} &=& \brb + \frac{L_{2}}{2} \tb  \bhub +  \frac{D}{2}  \nonumber \\ 
&& \times  \left \{  \cos ( \qb \tb  + \azb ) \bv + \sin ( \qb \tb + \azb )\bw_{2}  \right \} \label{dseg}
\end{eqnarray}
in terms of the dimensionless contour variables $t_{i}$ ($ -1 \le
t_{i} \le 1$) and coil pitch $k_{i} = 2\pi L_{i}/ p_{\text{int}}^{(i)}
$ with $p_{\text{int}}^{(i)}$ the distance along the longitudinal rod
axis over which the Yukawa helix makes a full revolution. With this convention  $k_i >0$ corresponds to a right-handed helix and $k_i < 0$ to a left-handed one.

\begin{figure}
\begin{center}
\includegraphics[clip=,width= 0.6 \columnwidth, angle = -90 ]{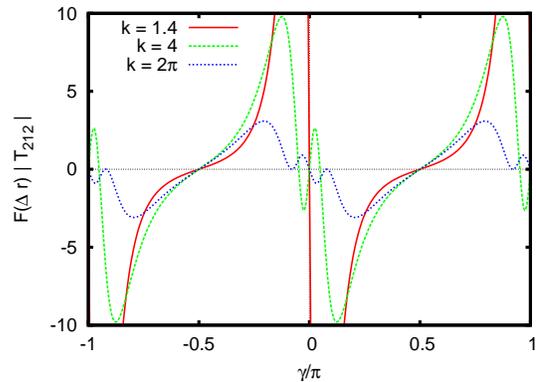}
\caption{ \label{twangle} Reduced chiral potential ${\mathcal
    F}(\Delta r)  |\chi|$ between two helical Yukawa rods as a function of the
  interrod angle $\gamma$ for fixed centre-of-mass distance $\Delta r =
  0.1L $ and  $\kappa L = 20$. }
\end{center}
\end{figure}

In this study we will focus on the general case of a binary mixture of
helical species with a {\em different} coil pitch (in terms of sign
and/or magnitude), i.e. $k_{1} \neq k_{2}$.  Since a helix is {\em
  not} invariant with respect to rotations about the longitudinal
pitch axis $\bhu_{i}$  we need to define an additional orientational unit vector ${\bf
  \hat{e}}_{i} \perp \bhu_{i}$  to account for its azimuthal degress
of freedom. Consequently, the interaction potential \eq{yukhel} must depend on a  set of {\em internal} azimuthal angles $\psi_{i}$ ($0 \le \psi_{i} \le 2 \pi$), defined such that  $\cos \psi_{i} = \bw_{i} \cdot {\bf \hat{e}}_{i}$. 
Using \eq{dseg} and employing orthogonality of the unit vectors allows
us to derive an expression for the norm of the segment-segment distance.
Let us normalize the centre-of-mass distance in units of rod
length of species 1, $L_{1}$, so that $\Delta \br \rightarrow \Delta
\br /L_{1}$. If we further assume the helices to be sufficiently slender
($D/L_{i} \ll 1$) we may  Taylor expand the potential up to leading
order in $D/L_{i}$:
\begin{widetext}
\begin{eqnarray}
\beta U_{Y} ( \Delta \br ; \bhua , \bhub ; \aza , \azb ) & \simeq & 
\beta U_{Y}^{(0)} ( \Delta \br ; \bhua , \bhub )  - Z_{1}Z_{2} \left (
  \frac{\lambda_{B}}{D} \right ) \left ( \frac{D}{L_{1}}  \right )^{2}
\int _{-1}^{1} d \ta \int _{-1}^{1} d \tb U^{\prime}_{Y} ( \Delta
\tilde{ r } ) \nonumber \\
&& \times  \left \{  \left [  \cos ( \qb \tb + \azb ) - \cos ( \qa \ta + \aza ) \right ] ( \bv \cdot \Delta \br ) \right . \nonumber \\
   && +    \left [ \sin (\qb \tb + \azb) ( \bw_{2} \cdot \Delta \br ) - \sin ( \qa \ta + \aza ) ( \bw_{1} \cdot \Delta \br ) \right ]  \nonumber \\
   && \left .  - (1/2) \left [ \ta \sin (\qb \tb + \azb) ( \bw_{2} \cdot \bhua ) + \ell \tb \sin ( \qa \ta + \aza ) ( \bw_{1} \cdot \bhub ) \right ] \right \} + {\cal O} [ (D/L)^{2} ] \label{usom} 
 \end{eqnarray}
\end{widetext}
in terms of the length ratio $\ell = L_{2}/L_{1}$ and  $U^{\prime}_{Y}$ the derivative of the Yukawa potential with respect to distance:
\beq
U^{\prime}_{Y} (\Delta \tilde{r}) = \frac{\exp [ -\kappa L_{1} \Delta \tilde{r} ]}{2 \Delta \tilde{r}} \left ( \frac{\kappa L_{1}}{\Delta \tilde{r}} + \frac{1}{\Delta \tilde{r}^{2}} \right ) \label{yukprime}
\eeq
and $\Delta \tilde{r}$ the distance between the positions $\ta$ and $\tb$ along the pitch axis of the helices (in units $L_{1}$):
\bea
 \Delta \tilde{r}^2  &=& \Delta r ^{2} +   \left [ \ell \tb ( \bhub
   \cdot \Delta \br) - \ta ( \bhua \cdot \Delta \br ) \right  ]  \nonumber \\ 
&& + \frac{1}{4} [ \ta ^2 + \ell^{2} \tb ^2 - 2 \ell \ta  \tb ( \bhua \cdot \bhub ) ] \label{rtilde}
\eea
The reference potential $\beta U_{Y} $ in \eq{usom} represents the pair potential between two screened line charges of length $L_{1}$ and $L_{2}$  interacting via the Yukawa potential \cite{kirch-bd}:
\bea 
\beta U_{Y}^{(0)} ( \Delta \br ; \bhua , \bhub)  & = & Z_{1} Z_{2} \left (
  \frac{ \lambda_{B}}{D} \right ) \left ( \frac{D}{L_1} \right )
\int_{-1}^{1} d \ta \int _{-1}^{1}  d \tb \nonumber \\
&& \times \frac{ \exp[- \kappa L_{1} \Delta \tilde{r} ]}{\Delta \tilde{r}} \label{yukline}
\eea
Since this potential is strictly achiral we will disregard it in the sequel of this paper. Focusing now on the second, chiral contribution in \eq{usom} and applying  standard trigonometric manipulations we  may recast it  into the following form:
\begin{eqnarray}
&& \beta U_{Y}^{(c)} ( \Delta \br ; \bhua , \bhub ; \aza , \azb )  = \nonumber \\ 
&&    - Z_{1}Z_{2} \left ( \frac{\lambda_{B}}{D} \right ) \left (
  \frac{D}{L_{1}}  \right )^{2} \int _{-1}^{1} d \ta \int _{-1}^{1} d \tb U^{\prime}_{Y} ( \Delta \tilde{ r } ) \nonumber \\
&& \times \left \{   {\cal A} \cos \aza + {\cal B} \cos \azb + {\cal C} \sin \aza + {\cal D} \sin \azb \right \} 
\end{eqnarray}
Henceforth, we will denote this potential by superscript ``$c$''. The coefficients are given by:
\begin{eqnarray}
{\cal A} & = & -  \cos (\qa \ta  )  (\bv \cdot \Delta \br) - \sin
(\qa \ta ) (\bw_{1} \cdot \Delta \br) \nonumber \\
&& - \frac{\ell \tb}{2}  \sin (\qa \ta ) (\bw_{1} \cdot \bhub ) \nonumber \\
{\cal B} & = &   \cos  (\qb \tb )  (\bv \cdot \Delta \br) + \sin  (\qb
\tb ) (\bw_{2} \cdot \Delta \br) \nonumber \\
&& - \frac{\ta}{2}  \sin  (\qb \tb ) (\bw_{2} \cdot \bhua )  \nonumber \\
{\cal C} & = & \sin  (\qa \ta )  (\bv \cdot \Delta \br) - \cos  ( \qa
\ta ) (\bw_{1} \cdot \Delta \br) \nonumber \\
&& - \frac{\ell \tb}{2}  \cos ( \qa \ta )(\bw_{1} \cdot \bhub )  \nonumber \\
{\cal D} & = & -  \sin ( \qb \tb )  (\bv \cdot \Delta \br) + \cos (\qb
\tb ) (\bw_{2} \cdot \Delta \br) \nonumber \\
&& - \frac{\ta}{2}  \cos ( \qb \tb ) (\bw_{2} \cdot \bhua )  \label{coef}
\end{eqnarray}
The next step is to construct an {\em angle-averaged} potential $\beta
{ \bar U_{Y}^{(c)}} $ by carrying out a proper preaveraging over the internal azimuthal angles. By requiring the Helmholtz free energy of the angle-averaged potential to be equal to that of the full angle-dependent potential  one can show that \cite{israelachvili}:
\begin{eqnarray}
 \beta { \bar U_{Y}^{(c)}}  & = &  - \ln  \left \langle   \exp [ -
   \beta U_{Y}^{(c)} ]  \right \rangle _{\psi_{1,2}} =    \left
   \langle    \beta U_{Y}^{(c)}    \right \rangle  _{\psi_{1,2}}
 \nonumber \\ 
&& -\frac{1}{2} \left ( \left \langle ( \beta U_{Y}^{(c)} )^{2}  \right \rangle _{\psi_{1,2}} - \left \langle  \beta U_{Y}^{(c)} \right \rangle  _{\psi_{1,2}} ^{2} \right ) + \cdots \label{angav} \nonumber \\
\end{eqnarray}
where the brackets denote a double integral over the internal angles
$\langle .\rangle _{\psi}  = (2 \pi)^{-1}\int_{0}^{2 \pi} d \psi
$. A similar expression can be obtained starting from a self-consistent
Boltzmann-weighted average of the chiral potential, i.e $\bar{U} = \langle
U\exp[-\beta U]\rangle_{\psi_{1,2}}/\langle \exp[-\beta U] \rangle
_{\psi_{1,2}} $ and Taylor expanding for small $\beta U$. This will give the
same result except for a trivial prefactor in the fluctuation term \cite{rushbrooke-1940,rowlinson-1958}.
The leading order contribution  $ \langle   \beta U_{Y}^{(c)}
 \rangle _{\psi_{1,2}}$ vanishes \cite{tombolatoferrarini} upon integrating over
$\psi_{i}$ so that we need to consider the next-order fluctuation
term in \eq{angav}. Using the isotropic averages $ \langle \cos ^{2} \psi \rangle _{\psi} =  \langle  \sin ^{2} \psi  \rangle _{\psi} = 1/2$ the angle-averaged chiral potential becomes:
\bea
\beta \bar{U}_{Y}^{(c)} ( \Delta \br ; \bhua , \bhub ) & \simeq &
-\frac{1}{4} Z_{1}^{2}Z_{2}^{2} \left ( \frac{\lambda_{B}}{D} \right
)^{2} \left ( \frac{D}{L_{1}}  \right )^{4}  \nonumber \\
&& \times \left (   {\cal \tilde{A}}^2  + {\cal \tilde{B}}^2  + {\cal \tilde{C}}^2 + {\cal \tilde{D}}^2  \right )
\eea
where
\beq
{\cal \tilde{A}} = \int _{-1}^{1} d \ta \int _{-1}^{1} d \tb U^{\prime}_{Y} ( \Delta \tilde{ r } ) {\cal A}
\eeq
and identical relations for the other coefficients.
A close inspection of the coefficients \eq{coef} reveals that only
those terms proportional to the pseudo-scalar $\bv \cdot \Delta \br$
contribute to the chiral potential. These are given by products
involving the first and third terms in \eq{coef}. All other
contributions are invariant under a parity transformation $\Delta \br
\rightarrow - \Delta \br$ which renders them irrelevant for the present analysis.
If we use the standard representation for the triple product $\bhua \times (\bhua \times \bhub) = \bhua (\bhua \cdot \bhub) - \bhub (\bhua \cdot \bhua)$ so that $ - \bw_{1} \cdot \bhub =  \bw_{2}\cdot \bhua = \sin \gamma  $ (with $\gamma$ the angle between the main axes of the helices) the chiral potential simplifies to:
\bea 
\beta \bar{U}_{Y}^{(c)} ( \Delta \br ; \bhua , \bhub ) & \simeq &
-\frac{1}{8} Z_{1}^{2}Z_{2}^{2} \left ( \frac{\lambda_{B}}{L_{1}}
\right )^{2} \left ( \frac{D}{L_{1}}  \right )^{2} \nonumber \\
&& \times {\cal F}_{12} (\Delta r ; k_{1} , k_{2} ) \chi\label{uchi}
\eea
in terms of a pseudo-scalar $\chi$ which changes sign under a parity
transformation \cite{meervertogen} $ \Delta \br \rightarrow - \Delta \br $:
\beq
\chi = (\bhua \times \bhub \cdot \Delta \hat{\br})
\eeq
with $\Delta \hat{\br} = \Delta \br /|\Delta \br |$ the centre-of-mass
unit vector. The function ${\cal F}_{12}$ depends on the
distance and helix orientations and is invariant under parity transformation:
\begin{eqnarray}
{\cal F}_{12} (\Delta r ; k_{1} , k_{2} ) & = &  - \langle \cos ( \qa
\ta  ) \rangle _{t_{1,2}} \langle  \ell \tb  \sin( \qa \ta  ) \rangle
_{t_{1,2}} \nonumber \\
&&  - \langle \cos ( \qb \tb  ) \rangle _{t_{1,2}} \langle  \ta  \sin( \qb \tb  ) \rangle _{t_{1,2}} \nonumber \\
&& +  \langle \sin ( \qa \ta  ) \rangle _{t_{1,2}} \langle \ell \tb
\cos( \qa \ta  ) \rangle _{t_{1,2}} \nonumber \\
&& + \langle \sin ( \qb \tb  ) \rangle _{t_{1,2}} \langle  \ta  \cos( \qb \tb  ) \rangle _{t_{1,2}} \label{ff}
\end{eqnarray}
where the brackets are short-hand notation for the double contour integration over $\beta U_{Y}^{\prime}$:
\beq
\langle  . \rangle _{t_{1,2}} = \Delta r ^{1/2}\int_{-1}^{1} d \ta \int_{-1}^{1} d \tb U_{Y}^{\prime}(\Delta \tilde{r}) \label{dcont}
\eeq
Since the prefactor ${\cal F}_{12}$ depends rather intricately on the
centre-of-mass distance and orientations it is desirable to seek a
simplified form. This can be achieved by ignoring the interactions
involving the ends of the helix. Indeed, if we  take the limit of
$L_{i}/D \rightarrow \infty $ the second contribution  in \eq{rtilde},
which embodies the interaction between the end of one rod with the
main section of the other, becomes vanishingly small. An equivalent
approach would be to fix the centre-of-mass distance vector along the unit vector $\Delta \br = \Delta r \bv $. In either case, the segment-segment distance \eq{rtilde} simplifies to:
\beq
 \Delta \tilde{r}^2  \simeq \Delta r ^{2} + \frac{1}{4} (\ta ^2 + \ell^{2} \tb ^2 - 2 \ell \ta  \tb \cos \gamma ) 
\eeq
and the only angular dependence is contained in the  angle $\gamma$
between the main axes of the helices. The $\gamma$-dependence of the
chiral potential at fixed
centre-of-mass distance has been plotted in \fig{twangle} and reveals
a strongly non-monotonic relation.
In a concentrated cholesteric nematic phase rods are usually strongly
aligned along the local director so that  $\gamma$ will on average
be very small. In the asymptotic limit of strong local orientational
order the following scaling expression for the chiral potential can be justified:
\beq
\beta \bar{U}_{Y}^{(c)} \propto \gamma {\cal F}_{12}(\Delta r ; k_1 , k_2 ) (\bv \cdot \Delta \hat{\br})
\eeq
where $ \Delta \tilde{r} = [\Delta r ^{2} + \frac{1}{4} (\ta  - \ell
\tb)^2 ]^{1/2}$.
Looking at the local minima appearing in \fig{twangle} it is obvious that
the asymptotic approximation has to be taken with some
caution in the dilute regime where the average $\gamma$ is no
longer very small. A striking anomaly occurs for $k=4$ where the
chiral torque
 $(\partial U_{Y}^{(c)}/\partial \gamma )_{\gamma = 0}$ has an opposite sign compared to the
other cases shown. Moreover, the local and global minimum occurring for $k=4$
correspond to {\em opposite} twist directions.  This hints to a subtle relationship between the
magnitude of the coil pitch and the sense of the
cholesteric director field which will be explored in detail further on in this study.
In the asymptotic approximation, the distance dependence of the
chiral potential is embodied by the chiral amplitude ${\cal F}_{12}$
which will be analyzed in the subsequent paragraphs where we will focus on identical
and enantiomeric helices, respectively.

\subsection{Identical helices}
 If  the helices  are identical, then $L_{1} = L_{2}=L$ ($\ell = 1$),  $k_{1} = k_{2} = k$ and $Z_{1}=Z_{2}=Z$. The double contour integration \eq{dcont} is invariant under interchanging $\ta \leftrightarrow \tb$ and we can exploit this symmetry to simplify \eq{ff} as follows:
\bea
{\cal F}  (\Delta r ;  k) &=&   - 2 \langle \cos ( k \ta  ) \rangle
_{t_{1,2}} \langle  \tb  \sin( k \ta  ) \rangle _{t_{1,2}} \nonumber \\ 
&& + 2 \langle \sin ( k \ta  ) \rangle _{t_{1,2}} \langle  \tb  \cos( k \ta  ) \rangle _{t_{1,2}}  
\eea
The second contribution involves odd terms in $\ta$ and $\tb$ which
vanish upon performing the double contour integration [\eq{dcont}]. The result is:
\bea
{\cal F}(\Delta r; k)  &=&   - 2 \Delta r \left ( \int_{-1}^{1} d \ta
  \int_{-1}^{1} d \tb U_{Y}^{\prime}(\Delta \tilde{r})\cos ( k \ta  )
\right ) \nonumber \\
&& \times    \left ( \int_{-1}^{1} d \ta \int_{-1}^{1} d \tb U_{Y}^{\prime}(\Delta \tilde{r}) \tb \sin ( k \ta  )  \right )    \label{fdr}
\eea
which still implicitly depends on the coil pitch $k$ and Debye screening length $\kappa L$ via \eq{yukprime}. 
The chiral potential between two identical helical Yukawa rods thus reads:
\beq
\beta \bar{U}_{Y}^{(c)} ( \Delta \br ; \bhua , \bhub )  \simeq    -\frac{1}{8} Z^{4} \left ( \frac{\lambda_{B}}{D} \right )^{2} \left ( \frac{D}{L}  \right )^{4} 
{\cal F} (\Delta r ; k) \chi \label{ucmono}
\eeq
\begin{figure}
\begin{center}
\includegraphics[clip=,width= 0.6 \columnwidth, angle = -90 ]{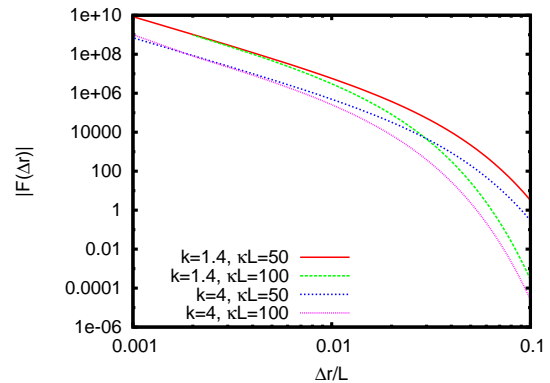}
\caption{ \label{eff} Amplitude of the chiral potential [\eq{fdr}]
  between two helical Yukawa rods as a function of centre-of-mass distance $\Delta r = \Delta \br \cdot \bv$.   }
\end{center}
\end{figure}
This potential, shown in \fig{eff}, is found to
decay steeply with increasing rod centre-of-mass distance. Owing to
the intractable double contour integrations, the distance-dependence
of the potential is essentially non-algebraic and does not obey a
simple power-law scaling with respect to $\Delta r$. At short
distances the amplitudes appear to be mainly dependent upon the coil pitch $k$ rather than the screening constant. 


It is worthwhile to compare the chiral potential \eq{ucmono} to the
one proposed by Goossens \cite{goossens,wensinkjackson} which has been
used frequently in literature. The Goossens potential emerges as the
leading order contribution from a multipolar expansion of the
interaction potential between two molecules each composed of an array
of dipoles. This potential takes the following generic form: 
\beq
U _{GS} ( \Delta \br ; \bhua , \bhub ) = -U_{0} \left (\frac{\sigma}{\Delta r} \right )^{7} (\bhua \cdot \bhub)  \chi \label{gooss}
\eeq
in terms of a reference size $\sigma$ and amplitude $U_{0}$ which
determines the handedness of the chiral interaction. As can be gleaned
from \fig{eff} the distance-dependence of the Goossens potential is much steeper compared
to the decay found from \eq{fdr}. We further remark that both chiral potentials are invariant with
respect to an inversion of the rod orientation ($\bhu_{i} \rightarrow
- \bhu_{i} $, or equivalently $\gamma \rightarrow \gamma \pm \pi$). 


\subsection{Enantiomers: helices with opposite handedness}

Let us now consider two helices of identical length $L_{1} = L_{2} = L$ and charge $Z_{1} = Z_{2}=Z$ but opposite pitch $k_{1}=  -k_{2}$. The corresponding cross interaction between a right-handed and left-handed helix  then follows from \eq{ff}:
\begin{eqnarray}
{\cal F}_{12} (\Delta r ; k_{1} , -k_{1}) & = &  - \langle \cos ( \qa
\ta  ) \rangle _{t_{1,2}} \langle  \tb  \sin( \qa \ta  ) \rangle
_{t_{1,2}}  \nonumber \\
&& - \langle \cos ( -\qa \tb  ) \rangle _{t_{1,2}} \langle  \ta  \sin( -\qa \tb  ) \rangle _{t_{1,2}}  \nonumber \\
&& +  \langle \sin ( \qa \ta  ) \rangle _{t_{1,2}} \langle \tb  \cos(
\qa \ta  ) \rangle _{t_{1,2}} \nonumber \\
&& + \langle \sin ( -\qa \tb  ) \rangle _{t_{1,2}} \langle  \ta  \cos( -\qa \tb  ) \rangle _{t_{1,2}} \nonumber \\
&  =  & 0
\end{eqnarray}
irrespective of the magnitude of the coil pitch. The implication
of this result is that an equimolar binary mixture of helices with equal shape but opposite handedness (a so-called `racemic' mixture) does not exhibit cholesteric nematic order. If the rods have different lengths ($\ell \neq 1$) they no longer  form an enantiomeric pair and the chiral interaction will generally be non-zero.

\section{Prediction of the cholesteric pitch}

In order to study the implications of the internal helical structure of the rods on the
(macroscopic) cholesteric order we need a statistical theory which is able to make a connection between the pair interaction of the particles and the local equilibrium orientational distribution $f(\bhu \cdot \bn)$ and cholesteric pitch.
Such a theory can be devised starting from the classical Onsager
theory \cite{onsager} for infinitely thin hard rods which exhibit
common uniaxial nematic order. The theory has been generalized by
Straley \cite{straleychiral,allenevans} for
aligned fluids with weakly non-uniform director fields, such as
a cholesteric liquid crystal, in which case elastic contributions must
be incorporated into the free energy. The Helmholtz free energy density $F/V$ of a binary mixture of slender chiral rods in a cholesteric phase of volume $V$ takes the following form:
\bea
\frac{\beta F}{V}  &=&  \rho  ( \ln \rho {\mathcal {\bar V}} - 1 ) +
\rho \sum_{i} x_{i} \int d \bhu f_{i} (\bhu ) \ln [ x_{i} 4 \pi f_{i}
( \bhu ) ] \nonumber \\
&& + \beta \sum_{i} \sum_{j} x_{i} x_{j} \left (  K_{0}^{ij}  - K_{1}^{ij} q + \frac{K_{2}^{ij}}{2} q^{2} \right )
\eea
In terms of the thermal energy $\beta^{-1} = k_{B}T $, overall
particle density $\rho = N/V$. ${\mathcal {\bar V}} =
\prod_{i}{\mathcal V}^{x_{i}}$ is a weighted product of the thermal
volume  ${\mathcal V}_{i}$ of particle $i$ which includes
contributions arising from the rotational momenta. The first two terms
in the free energy denote the ideal translational and orientational
entropy of the system while the last one represents the excess free
energy which accounts for the interactions between the rods on the approximate second-virial level. The latter consists of three contributions. The first, $K_{0}$, involves a spatial and orientational average of the Mayer function $f_{M}=\exp[-\beta U]-1$ weighted by the orientational distribution functions (ODF) $f_{i}(\bhu)$ of the respective species:
\bea
\beta K_{0}^{ij} &=& - \frac{\rho^2}{2}\int d \bhua   f_{i}(\bhua) \int
d \bhub f_{j} (\bhub) \nonumber \\ 
&& \times \int d \Delta \br  f^{ij}_{M} (\Delta \br ; \bhua , \bhub ) \label{k0}
\eea
For hard anisometric bodies, the spatial integration leads to the {\em excluded volume} $v_{\text{excl}}^{ij}(\bhua, \bhub)$ between particles of species $i$ and $j$.
The second and third contributions in the excess free energy represent
the change of free energy due to the twist deformation of the director
field. The strength of this deformation is measured by the cholesteric
pitch $q = 2\pi/ p $, with $p \gg L \gg p_{\text{int}}$ the
 pitch distance associated with the helical director
field. Since the theory is only valid for long-wavelength distortions
of the director field it is required that $q \ll 1 $.  The
torque-field contribution, proportional to $K^{ij}_{1}q$ arises from
the chiral torque imparted by the chirality of the particles and leads
to a reduction of the free energy. Opposing this, there is an elastic
response counteracting the deformation of the director field. The
corresponding free energy penalty $K^{ij}_{2}q^2$ is proportional to
the {\em twist elastic constant} $K_{2}^{ij}$. Similar to $K_{0}$, the
torque-field  and twist elastic constants are given by spatio-angular
averages of the Mayer function, albeit in a more complicated way \cite{allenevans}:  
\begin{eqnarray}
\beta K_{1}^{ij} & = & - \frac{\rho^2}{2} \int d \bhua   f_{i}(\bhua)
\int d \bhub \dot{f}_{j} (\bhub) (\bhub \cdot \by ) \nonumber \\
&& \times  \int d \Delta \br (\Delta z) f^{ij}_{M} (\Delta \br ; \bhua , \bhub )  \nonumber \\
\beta K_{2}^{ij} & = & - \frac{\rho^2}{2} \int d \bhua
\dot{f}_{i}(\bhua) (\bhua \cdot \by ) \int d \bhub \dot{f}_{j} (\bhub)
(\bhub \cdot \by) \nonumber \\
&& \times \int d \Delta \br (\Delta z)^{2} f^{ij}_{M} (\Delta \br ; \bhua , \bhub )  \label{twistk}
\end{eqnarray}
where $\dot{f}$ represents the derivative of the ODF with respect to its argument. In arriving at \eq{twistk}  we have implicitly fixed the pitch direction along the $z$-direction of the laboratory frame with the local nematic director $\bn(z)$ pointing parallel to the  $x$-axis. 

Let us now consider a model binary mixture of hard rods with different
lengths ($\ell \neq 1$) decorated with a weak chiral potential of the
form proposed in the previous section [\eq{uchi}]. The twist elastic
constant is not affected by the weak chiral potential, but only
by the achiral hard cylindrical backbone and electrostatic reference
potential. If we neglect the latter contributions for the time being \footnote{The effect of the Yukawa reference potential could in principle be taken into account by introducing an {\em effective} hard-core diameter $D_{\text{eff}}>D$ which depends on the range of the electrostatic potential.} the spatial integrals appearing in \eq{k0} and \eq{twistk}  reduce to integrals over the excluded volume manifold of two thin hard cylinders weighted over powers of the pitch distance variable $\Delta z$:
\beq
M_{n}^{ij} (\bhua , \bhub) = \int _{\in v^{ij}_{\text{excl}}} \Delta \br  ( \Delta z )^{n}, \hspace{0.5cm} n=0,2
\eeq
These quantities have been calculated in general form for hard spherocylinders in \olcite{wensinkjackson}. Here, we only need the leading order contributions for large $L_{i}/D$:
\begin{eqnarray}
M^{ij}_{0} (\bhua , \bhub) & = &   v^{ij}_{\text{excl}}(\bhua, \bhub) = 2 L_{i}L_{j}D |\sin \gamma |  \nonumber \\
M^{ij}_{2} (\bhua , \bhub) & = & \frac{2}{3} L_{i} L_{j} D | \sin \gamma | \left ( \frac{L_{i}^{2}}{4} (\bhua \cdot \bz)^{2 } \right .\nonumber \\
&& \left . + \frac{L_{j}^{2}}{4} (\bhub \cdot \bz)^{2 } + D^{2} (\bv \cdot \bz)^{2} \right )  \label{mk}
\end{eqnarray}
Due to symmetry reasons the torque-field constant $K_{1}$ depends
only the chiral part of the potential and not on the achiral
reference part. For weak chiral potentials considered here it is
justified to approximate
$f_{M} \approx  -\beta U $, analogous to a van der
Waals perturbation approximation generalized to liquid crystals \cite{gelbartbaron,francomelgar,wensinkjackson}.
The spatial integration pertaining to $M_{1}$ then becomes:
\beq
M_{1}^{ij} (\bhua , \bhub) = \int _{\notin v^{ij}_{\text{excl}}} \Delta \br  ( \Delta z ) \beta {\bar U}_{Y}^{(c)}(\Delta \br ; \bhua , \bhub )
\eeq
where the spatial integral runs over the space complementary to the excluded volume of the particles. 
By exploiting the cylindrical symmetry of the cholesteric system one can parametrize the distance vector in terms of cylindrical coordinates so that $\Delta \br = r \sin \zeta \bx + r \cos \zeta \by + \Delta z \bz$. With this, one can write \footnote{Strictly, this expression is only valid if both rods are oriented perpendicular to the pitch axis $\bz$. This approximation can also be justified in case the local nematic order is asymptotically strong.}:
\beq
M_{1}^{ij} (\bhua , \bhub) = 
\frac{1}{8}Z_{i}^{2}Z_{j}^{2}\lambda_{B}^{2} D^{2} {\mathcal W}(k_i, k_j) (\bhua \times \bhub \cdot \bz) \label{m1f}
\eeq
where ${\mathcal W}$ represents a spatial integral over the
amplitude of the chiral potential for a given pitch $k_{i}$ of the
Yukawa coil of species $i$:
\beq
{\mathcal W}(k_i , k_j) =  -4 \pi \int_{0}^{\infty} d  r
r\int_{D/L_1}^{\infty}d \Delta z (\Delta z)^{2} {\cal F}_{ij}(\Delta
r; k_i , k_j)  \label{wee}
\eeq
with  $\Delta r = (r^2 + \Delta z^{2})^{1/2}$ the centre-of-mass distance parametrized in terms of cylindrical coordinates $\Delta r$ and $r$ (both in units of $L_{1}$). 

The next step is to perform a  double orientational averages of the
moment contributions $M_{k}$ according to \eq{twistk}. It is expedient
to adopt the Gaussian  approximation, in which the ODF is represented
by  $f_{i}(\bhu_i) \propto \exp[-\alpha_{i} (\bhu_{i} \cdot
\bn)^{2}/2]$ in terms of a single variational parameter $\alpha_{i}$ whose equilibrium value is required to minimize the total free energy. If the local nematic order is strong enough ($\alpha_{i} \gg 1$) the orientational averages can be estimated analytically  by means of an asymptotic expansion for small inter-rod angles. This procedure has been outlined in detail in \olcites{OdijkLekkerkerker,wensinkjackson}. The result for the nematic reference contribution $K_{0}^{ij}$ reads (up to leading order in $\alpha_{i}$):
\bea
\beta K_{0}^{ij} & \sim & \rho^{2} L_{i} L_{j} D \lang \gamma \rang \nonumber \\
 & \sim & \rho^{2} L_{i} L_{j} D \left ( \frac{\pi}{2} \right )^{1/2} \left ( \frac{\alpha_{i} + \alpha_{j}}{\alpha_{i} \alpha_{j}} \right )^{1/2}
\eea 
where the double brackets denote Gaussian orientational averages, specified in the Appendix. Similarly, one can derive for the twist elastic constant:
\begin{widetext}
\bea
\beta K_{2}^{ij} & \sim &  - \frac{ \rho^{2} }{192} L_{i} L_{j} D \alpha_i \alpha_j \left [ L_{i}^{2} \left  ( \lang \gamma \theta_{i}^{2} ( \theta_{i}^{2} + \theta_{j}^{2} ) \rang  - \lang \gamma^{3} \theta_{i}^{2} \rang \right )    +  L_{j}^{2} \left  ( \lang \gamma \theta _{j}^{2} ( \theta _{i}^{2} + \theta _{j}^{2} ) \rang - \lang \gamma ^{3} \theta _{j}^{2}  \rang \right ) \right ] \\ \nonumber
&& \sim \frac{\rho^2}{192} (2 \pi)^{1/2} L_{i} L_{j} D \left ( \frac{L_{i}^{2}(3 \alpha_{j}^{2} + 4 \alpha_{i}\alpha_{j}) + L_{j}^{2}(3 \alpha_{i}^{2} + 4 \alpha_{i}\alpha_{j})}{\alpha_{i}^{1/2} \alpha_{j}^{1/2} (\alpha_{i} + \alpha_{j})^{3/2}} \right )
\eea
\end{widetext}
Here we have ignored the last term in  $M_{2}^{ij}$ [\eq{mk}] which is negligible for slender rods. Finally, the torque-field contribution reads:
\beq
\beta K_{1}^{ij} \sim   \frac{\rho^2}{8} Z_{i}^{2} Z_{j}^{2}
\lambda_{B}^{2} D^{2} {\mathcal W}(k) \label{tf}
\eeq
With this, the free energy is fully specified. Minimization with respect to  $q$ yields for the equilibrium pitch:
\beq
q = \frac{ \sum_{i} \sum_{j} x_{i} x_{j} \beta K_{1}^{ij}}{ \sum_{i} \sum_{j} x_{i} x_{j} \beta K_{2}^{ij}}
\eeq
For weak pitches ($q \ll 1$), the local nematic order is only slightly
affected by the twist director field and the equilibrium values for
$\alpha_{i}$ depend only on the free energy of the nematic reference
phase ($q=0$). Minimization with respect to $\alpha_{i}$ and rearranging terms leads to the following set of coupled equations:
\bea
\alpha_{1} & = & \frac{\pi c^{2}}{4} \left ( x_{1} + 2^{1/2} x_{2} \ell  (1+ Q^{-1} )^{-1/2} \right )^{2} \nonumber \\
Q & = & \left ( \frac{2^{1/2} x_{1} \ell (1 + Q )^{-1/2} + x_{2} \ell^{2} }{ x_{1} + 2^{1/2} x_{2} \ell (1 + Q^{-1})^{-1/2}} \right ) ^{2}
\eea
in terms of the ratio  $Q = \alpha_{2}/\alpha_{1}$, dimensionless concentration $c = NL_{1}^{2}D/V$ and length ratio $\ell = L_{2}/L_{1}$. These equations cannot be solved analytically but the solutions are easily obtained by iteration. It is important to note  that both  $\alpha_{1}$ and $\alpha_{2}$ increase quadratically with the concentration $c$ since their ratio $Q$ only depends on the mole fractions $x_{i}$. 

\subsection{Monodisperse systems}

\begin{figure}
\begin{center}
\includegraphics[clip=,width= 0.6 \columnwidth, angle = -90 ]{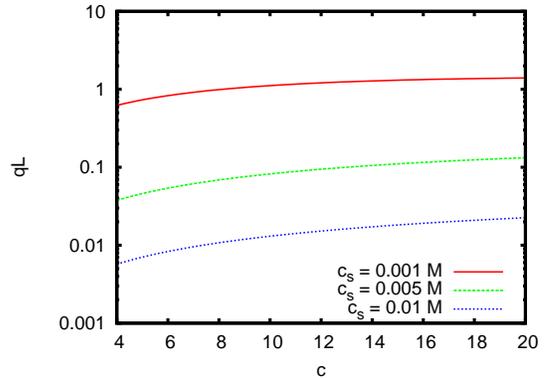}
\caption{ \label{pitchc} Variation of the cholesteric pitch $qL$ with concentration $c$ for right-handed Yukawa helices with $k=1.4$ and $L/D=50$.  The charge density is $\sigma_{c}D =1$ and the reduced Bjerrum length $\lambda_{B}/D = 0.1$.}
\end{center}
\end{figure}

For pure systems of infinitely thin helices, the twist elastic
constant behaves asymptotically as:
\beq
\beta K_{2} D \sim \frac{14 c}{192}
\eeq
as found by Odijk \cite{odijkchiral}. For the cholesteric pitch we obtain:
\beq
qL = \frac{\beta K_{1}DL}{\beta K_{2}D} \sim \frac{12}{7} c {\mathcal E}_{c} \label{pmono}
\eeq
where  ${\mathcal E}_{c}(\Delta r)$ is a dimensionless parameter
pertaining to the chiral potential between rods. It combines the microscopic characteristics of the helical rod such as the aspect ratio, surface charge and coil pitch [{\em viz.} \eq{m1f}]:
\beq
{\mathcal E}_{c}  \sim    \left (  \frac{\lambda_{B}}{D} \right )^{2}
Z^{4} \left ( \frac{D}{L} \right )^{3} {\mathcal W}(k)   \label{emono}
\eeq
Since the electrostatic screening of the charge-mediated chiral interactions depends on the rod concentration the variation of the pitch with concentration is strictly non-linear, as shown in \fig{pitchc}. 
To simplify matters we may state that typically $\lambda_{B}/D \sim {\mathcal O}(10^{-1})$
for rod-like colloids in water. The rod charge  $Z$ is expected to be
linearly proportional to the rod length $L$. Let us further introduce a charge
density $\sigma_{c}$, defined as the number of unit charges per unit
length, so that  $ Z \sim \sigma_{c} L$ and  $Z^{4} (D/L)^{3} \sim
{\mathcal O}(L/D)$ (assuming the charge density $\sigma_{c}D$ to be of
order unity). We remark that the $qL$ values shown in \fig{pitchc} correspond to  pitch
distances of 10-100 rod lengths, a range which  is commonly found in experiments \cite{dogic-fraden_chol,strey-dna}.

 \fig{pitchc} shows that the pitch is a monotonically increasing 
function of the rod concentration with the  magnitude depending strongly on the salt concentration. 
At high $c_{\text{s}}$ the electrostatic repulsion between the
Yukawa sites on the rods is strongly screened and the resulting chiral
interaction will be attenuated significantly. The screening effect is common in dispersions of
{\em fd} and DNA and supports the idea that chiral interactions are
primarily transmitted via long-range electrostatic interactions
\cite{grelet-fraden_chol,tombolato-grelet}. 
We remark that for short-fragment (146 base pair) DNA the opposite
trend is observed \cite{strey-dna}. There, the cholesteric pitch is
found to {\em increase} with respect to salt concentration, i.e.  charge screening leads to stronger
chiral interactions between the DNA chains.  This trend is most
likely explained by the steric effect associated with the double helical backbone which
becomes more pronounced as the charged phosphate groups residing on
the backbone become increasingly screened. Clearly, the excluded volume
of the helical grooves (neglected in this study) must be included
explicitly to give a proper account of both steric and electrostatic contributions to microscopic chirality in DNA \cite{tombolatoferrarini}.  

We remark that the effect of the temperature on
the chiral strength is rather trivial in our model.  At high temperatures the chiral
dispersion potential will become increasingly less
important than the achiral hard-core potential associated with the
cylindricasl backbone and a reduction of the pitch
is expected. This effect has been observed in {\em fd}
virus rods \cite{dogic-fraden_chol,tombolato-grelet}.

\begin{figure}
\begin{center}
\includegraphics[clip=,width= 0.6 \columnwidth, angle = -90 ]{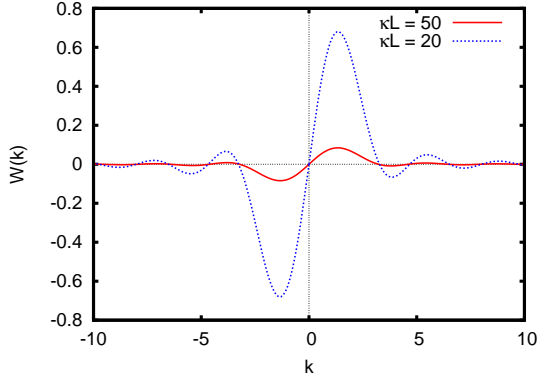}
\caption{ \label{ww} Spatially integrated chiral potential ${\mathcal
    W}(k)$ [\eq{wee}] versus coil pitch $k=2 \pi L/p_{\text{int}}$
  for identical  Yukawa helices with a right-handed ($k>0$) or
  left-handed ($k<0$) symmetry for $L/D = 50$. For ${\mathcal W} > 0$ a right-handed
  cholesteric phase is formed, whereas ${\mathcal W} <0 $ corresponds
  to a left-handed one. Zero-points indicate a sense inversion of the
  cholesteric helix at $k=3.25,4.67,6.51,8.00,..$.  }\end{center}
\end{figure}

The symmetry of the chiral interaction is entirely governed by the
spatial integral over the chiral potential which depends rather
intricately on the coil pitch of the  Yukawa helix and the Debye
screening length.  \fig{ww} shows the typical behavior of the
intrinsic chiral strength for two helical rods with $L/D=50$ as a
function of the coil pitch. For small coil pitches,
i.e. weakly coiled helical rods, the handedness of helical director
field is commensurate with that of the Yukawa coil, e.g 
right-handed helices form a right-handed cholesteric phase. The maximum twisting
effect is obtained for $k = 1.35 $. This value corresponds to
a coil pitch distance of $p_{\text{int}} \sim 5L$
which implies that only a marginal degree of coiling is required to
minimize the cholesteric pitch. 

A further increase of $k$ leads to a reduction of the chiral strength
and a sign change at $k=3.26$ corresponding  to a {\em sense inversion} of cholesteric helix. For
$|k| > 3.26$ the handedness of the cholesteric helix is no longer
commensurate with that of the Yukawa helix so a left-handed
cholesteric state is obtained from right-handed helices and vice
versa. Increasing $k$ even further reveals a  oscillatory relation
between the  microscopic (coil) and macroscopic (cholesteric) pitches. The
inversion points depend primarily on the coil pitch and
are not affected by the electrostatic screening $\kappa D$. At high $k$, the twisting
potential  strongly decays and vanishes asymptotically for tightly
coiled Yukawa rods ($k \rightarrow \infty$) where the effective `width' of the helical grooves becomes negligibly small.

\subsection{Binary mixtures of helical rods with equal lengths ($\ell = 1$)}

For a mixture of rods with equal length but different coil pitches ($k_{1} \neq k_{2}$) and/or surfaces charges ($Z_{1} \neq Z_{2}$), the situation is comparable to the monodisperse case since the twist elastic constant is independent of these properties. Furthermore, we have $\alpha_{1} = \alpha_{2} = \alpha $ and the cholesteric pitch is given by a form analogous to \eq{pmono}:
\beq
qL  \sim \frac{12c}{7} {\bar {\mathcal E}}_{c} \label{peqlength}
\eeq
in terms of an {\em effective} chiral strength given by a simple mole fraction average of the different pair contributions:
\beq
 {\bar {\mathcal E}}_{c} =  \sum_{i} \sum_{j} x_{i} x_{j} {\mathcal E}_{c}^{ij} \label{epseff}
\eeq
with  ${\mathcal E}_{c}^{ij} $ the generalized version of \eq{emono}:
\beq
{\mathcal E}_{c}^{ij}  \sim    \left (  \frac{\lambda_{B}}{D} \right
)^{2} Z_{i}^{2}Z_{j}^{2} \left ( \frac{D}{L_{1}} \right )^{3}{\mathcal W}(k_{i}, k_{j})   \label{ebi}
\eeq

\subsection{Binary mixtures of helical rods with unequal lengths ($\ell \neq 1$)}

In case of a binary mixture of rods with different lengths things are more complicated and most thermodynamic properties such as the cholesteric pitch can only be assessed numerically. Let us first investigate the effect of a weak chiral potential on the phase behavior of rod mixtures with different length ratios. To that end we must first consider the free energy of a binary mixture of infinitely thin hard rods with length ratio $\ell$ in the nematic phase. Within the Gaussian approximation it is given by \cite{OdijkLekkerkerker,Vroege&Lekkerkerker}:
\bea
\frac{\beta F_{\text{nem}}}{N} &  \sim  & 
\ln c  + \sum_{i} x_{i} [ \ln x_{i} \alpha_{i}   - 1] \nonumber \\ 
&& + c \sum _{i} \sum_{j} x_{i} x_{j} \frac{L_{i}L_{j}}{L_{1}^{2}} \lang \gamma \rang \nonumber \\
& \sim & \ln c  + \sum_{i} x_{i} [ \ln x_{i} \alpha_{i}   - 1] + c
\left (  \frac{\pi}{\alpha_{1}} \right )^{1/2} \nonumber \\ 
&& \times  \left [ x_{1}^{2} + 2^{1/2} x_{1} x_{2} \ell (1+Q^{-1})^{1/2} + x_{2}^{2} \ell^{2} Q^{-1/2} \right ] 
\eea
In the isotropic phase the free energy simplifies to:
\beq
\frac{\beta F_{\text{iso}}}{N}  \sim   \ln c  + \sum_{i} x_{i} \ln x_{i} + c \frac{\pi}{4} \left [ x_{1}^{2} + 2 x_{1}x_{2} \ell + x_{2}^{2} \ell^{2} \right ] \\
\eeq
In the cholesteric phase we must take into account the change of free energy  associated with the weak helical distortion of director-field as discussed in the preceding Section:
\beq
\frac{\beta F_{\text{chol}}}{N}  \sim  \frac{\beta F_{\text{nem}}}{N}  + \frac{1}{c} \sum_{i} \sum_{j} x_{i}x_{j} \left [  -\tilde{K}_{1}^{ij} \tilde{q} + \frac{1}{2} \tilde{K}_{2}^{ij} \tilde{q}^{2} \right ] 
\eeq
which yields upon minimization with respect to the cholesteric  pitch $\tilde{q}  = 2 \pi L_{1}/p$:
\beq
\frac{\beta F_{\text{chol}}}{N}  \sim  \frac{\beta F_{\text{nem}}}{N}  - \frac{1}{2c} 
  \frac{ \left (  \sum_{i} \sum_{j} x_{i}x_{j} \tilde{K}_{1}^{ij}  \right )^{2}}{\sum_{i} \sum_{j} x_{i}x_{j} \tilde{K}_{2}^{ij}}
\eeq
We reiterate that this expression is only applicable in the regime $\tilde{q} \ll 1$ where the helical distortion of the director field does not interfere with the local orientational order. Since the twist elastic constants are positive, the second term in the free energy must be negative which implies that a small degree of chirality {\em always} leads to a reduction of the free energy of the system.  In explicit form, the twist elastic constants for the pure components are given by:
\bea
\tilde{K}_{2}^{11} & = & \beta K_{2}^{11}D \sim  \frac{7c^2}{192} \left ( \frac{\pi}{\alpha_{1}} \right )^{1/2} \nonumber \\
\tilde{K}_{2}^{22} & = & \beta K_{2}^{22}D \sim  \frac{7c^2}{192}  \left ( \frac{\pi}{\alpha_{1}} \right )^{1/2} \ell^{4} Q^{-1/2} \nonumber \\
\tilde{K}_{2}^{12} & = & \beta K_{2}^{12}D \sim  \frac{7c^2}{192}  \left ( \frac{\pi}{\alpha_{1}} \right )^{1/2} g(Q)
\eea
with
\beq
g(Q) =   2^{1/2} \frac{(3Q^{2} + 4Q)\ell + (3 + 4Q) \ell^{3}}{7Q^{1/2} (1+Q)^{3/2}}
\eeq
Similarly, one can produce for the torque-field contributions:
\beq
\tilde{K}_{1}^{ij}  =  \beta K_{1}^{ij}DL_{1} \sim \frac{c^2}{8}  {\mathcal E}_{c}^{ij} \nonumber \\
\eeq
With this, the cholesteric free energy of a binary mixture with $\ell \neq 1$ can be rewritten as:
\beq
\frac{\beta F_{\text{chol}}}{N}  \sim  \frac{\beta F_{\text{nem}}}{N}  - \frac{3c^{2}}{28} G(x_{i},Q) {\bar {\mathcal E}}_{c}^{2}
\eeq
with:
\beq
G(x_{i},Q) = \left ( \frac{x_{1} + 2^{1/2}x_{2} \ell (1+Q^{-1})^{-1/2}}{x_{1}^{2} + 2x_{1}x_{2}g(Q) + x_{2}^{2} \ell^{4} Q^{-1/2} } \right )
\eeq
The corresponding cholesteric pitch takes the following form:
\beq
\tilde{q} = \frac{2 \pi L_{1}}{p} \sim \frac{12c}{7} {\bar {\mathcal E}}_{c} G(x_{i},Q)
\eeq
As required, this expression reduces to the forms proposed in the
previous subsections [\eq{pmono} and \eq{peqlength}] if we substitute
$\ell =1$ and $Q=1$. Recalling that $Q=Q(x)$ we conclude that the
concentration dependence of the cholesteric pitch of the mixture is
{\em identical} to that of the monodisperse system. The behavior of
the pitch with respect to the mole fraction $x_{2} = 1 - x_{1}$ on the
other hand is expected to be quite rich and this will be scrutinized in detail further on.

Let us now consider a binary mixtures of helical rods with equal charge density $\sigma_{c}$ so that the mole-fraction averaged chiral parameter ${\bar {\mathcal E}}_{c}$ can be written in compact form as:
\beq
{\bar {\mathcal E}}_{c} = \varepsilon_{0} \left ( x_{1}^{2}  \varepsilon_{11} + 2 x_{1}x_{2} \ell^{2} \varepsilon_{12} + x_{2}^{2} \ell^{4} \varepsilon_{22} \right )
\eeq
in terms of an amplitude $\varepsilon_0$ and phase factor $\varepsilon_{ij}$:
\bea
\varepsilon_{0} & = &  \left (  \frac{\lambda_{B}}{D} \right )^{2} \left (  \frac{L_{1}}{D} \right )  (\sigma_{c}D)^{4} \nonumber \\
\varepsilon_{ij}  & = &  {\mathcal W}(k_{i} , k_{j})
\eea
which needs to be specified for the mixture of interest. 

\begin{figure*}
\begin{center}
\includegraphics[clip=,width= 0.4 \columnwidth, angle = -90 ]{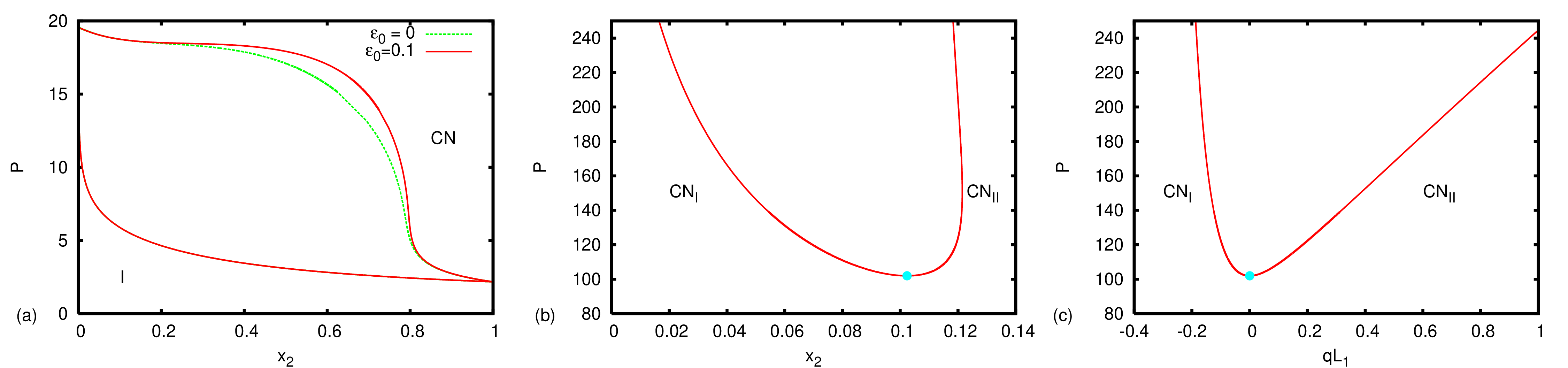}
\includegraphics[clip=,width= 0.4 \columnwidth, angle = -90 ]{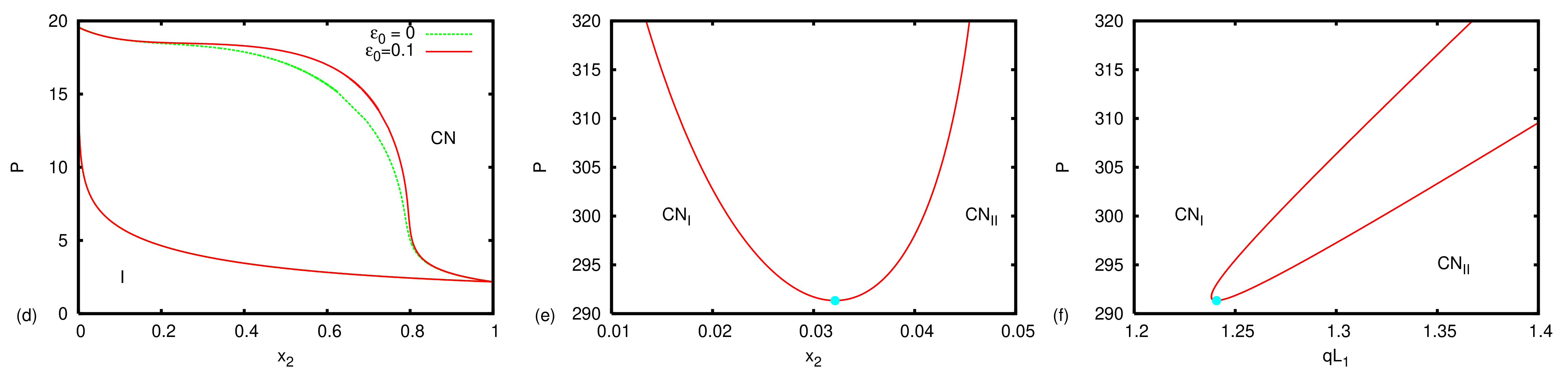}
\caption{ \label{phase} (a) Isotropic-cholesteric nematic phase
  diagram for a mixture of Yukawa helices with length ratio $\ell = 3$
  consisting of helices with opposite handedness ($k_{1}= -k_{2} =
  1.4$). Plotted is the osmotic pressure $P = \beta \Pi L_{1}^{2}D$
  versus the mole fraction of the left-handed species $x=x_{2}$. (b)
  High pressure region featuring a demixing of the cholesteric
  phase. A critical point is indicated by the dot. (c) Cholesteric
  pitch of coexisting left- and right-handed cholesteric phases versus pressure. (d)-(f) Same results for a mixture of helices with equal handedness ($k_{1}=k_{2}=1.4$) and $\ell = 3$.}
\end{center}
\end{figure*}

Let us first consider the case $\ell=3$ with $k_{1} = -k_{2} = 1.4$,
i.e. a  mixture of short right-handed helices mixed with long
left-handed ones with equal pitch magnitude. To simplify matters we
shall consider dispersions with excess added salt such that the
screening constant is fixed to  $\kappa L_{1} = 50$, independent of
the rod concentration. The phase diagrams of the chiral systems and
the corresponding hard rod reference system are shown in
\fig{phase}. Althought the isotropic-nematic transition is only marginally
affected by chirality, the cholesteric phase show a demixing into two coexisting cholesteric fractions with
opposite handedness. The demixing region is closed off by a lower
critical point located at $qL=0$ indicating that the cholesteric pitch
vanishes at the critical point. Note that the achiral mixture do not
show this demixing at this particular length
ratio. Within the Gaussian approximation \cite{Vroege&Lekkerkerker}
such a demixing  only occurs above a critical length ratio  $\ell >
3.167$. Moreover, the nematic-nematic  binodals do not meet
in a lower critical point but  merge with the
isotropic-nematic ones to produce an isotropic-nematic-nematic triphasic
equilibrium, irrespective of the length ratio $\ell$. The
cholesteric-cholesteric demixing is driven primarily  by the small chiral
dispersion contribution to the rod interaction potential and does not
arise from an interplay of the various entropic contributions (associated
with mixing, free-volume and orientational order) such as for hard rod mixtures. A similar demixing is observed for a mixtures of
right-handed helices with $k_1 = k_2 = 1.4$ at $\ell =3$ and amplitude
$\varepsilon_{0} = 0.1$,  albeit at a much higher osmotic pressure. It
is worth noting that for this case the cholesteric pitch does not reduce to zero at the critical point but attains a finite positive value.
For the case $\ell =1$, i.e. helices with equal lengths, no demixing
of the cholesteric state was found. This shows that a cholesteric mixture of left-
and righthanded helices can only demix if the length ratio between the
species is considerable.   

\begin{figure}
\begin{center}
\includegraphics[clip=,width= 0.7 \columnwidth, angle = -90 ]{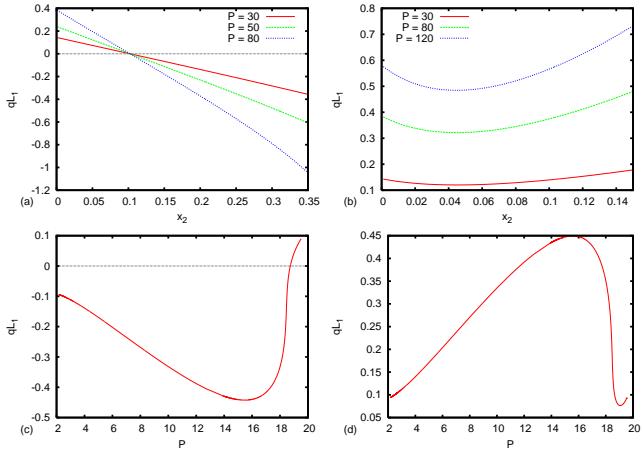}
\caption{ \label{px} Variation of the cholesteric pitch versus mole fraction at constant pressure. (a) Helices with opposite handedness ($k_{1}=-k_{2}=1.4$, $\ell = 3$). (b) Helices with equal handedness ($k_{1} = k_{2} = 1.4$, $\ell =3$). (c) Behavior of the pitch as a function of the isotropic-cholesteric coexistence pressure for $k_{1}=-k_{2}=1.4$ and $\ell = 3$. (d) Same for $k_{1} = k_{2} = 1.4$, $\ell =3$.}
\end{center}
\end{figure}

\fig{px} shows the behavior of the cholesteric pitch across the range
of mole fractions.  For the first  mixture, a quasi-linear reduction of
the pitch is observed accompanied by a change of handedness upon
increasing the mole fraction of the left-handed `dopant' helices. The
zero-point corresponding to a vanishing pitch is  virtually (but not
strictly) independent of the osmotic pressure of the suspension. For
the second mixture the trend is completely different and features a
non-monotonic increase of the pitch with mole fraction.  A similar
non-monotonic trend is  observed for the pitch of a cholesteric phase in
coexistence with an isotropic phase as a function of pressure.

\section{Discussion and conclusions}

It is instructive to compare the present Yukawa-type chiral potential with a much simpler model potential employed in a previous study \olcite{wensinkjackson}. There, we have used a {\em square well} (SW) chiral potential based on the Goossens form \eq{gooss}: 
\beq
U _{\text{SW}} ( \Delta \br ; \bhua , \bhub ) = -\varepsilon_{\text{SW}} H(\Delta r - \lambda)(\bhua \cdot \bhub)  \chi \label{sw}
\eeq
with $H(\Delta r - \lambda)$ a Heaviside step function. The main parameters characterizing the chiral interaction are the SW range  $\lambda >(1+D/L)$ and depth $\varepsilon_{\text{SW}}$. The main advantage of using the SW form is it that it renders the spatial integrations over the potential analytically tractable. The result for the torque field constant is very simple: $\beta K_{1}D^{2} \sim (\pi c^{2}/6) \beta \varepsilon_{\text{SW}}  \lambda^{4}$. Comparing this with \eq{tf} and setting the reduced SW range $\lambda $ to unity allows us to make an explicit link between the SW depth and the microscopic parameters pertaining to the electrostatic interactions between the helices:
\beq
\beta \varepsilon_{\text{SW}} \sim \frac{3Z^{4}}{4\pi} \left (
  \frac{\lambda_{B}}{D} \right )^{2} \left( \frac{D}{L} \right )^{4}{\mathcal W}(k) 
\eeq
The justification for this relation lies in the notion that most
thermodynamic properties are governed by the {\em spatially integrated pair potential} rather than the bare one. A prominent example is the
classical van der Waals model for fluids whose universal nature stems from
the fact that any arbitrary (but weakly) attractive potential can be mapped onto a single
integrated van der Waals energy which, along with the excluded-volume
contribution, fully determines the equation of state and hence the
thermodynamics of the fluid state.

In summary, we have proposed a helical Yukawa segment model as
a course-grained model in an effort to quantify the chiral interaction between
stiff chiral polyelectrolytes. Our chiral potential strongly resembles the classic Goossens potential
\cite{goossens}, which has been used almost exclusively to describe
long-ranged chiral dispersion forces. Contrary to the Goossens form, our potential provides an explicit reference
to the microscopic and electrostatic properties of the rods.
Combining the potential
with a simple second-virial theory allows us to study the structure and thermodynamic
stability of the cholesteric state as a function of rod density,
degree of coiling, and electrostatic screening. While the magnitude of
the cholesteric pitch depends
strongly on the rod concentration and concentration of added salt, the
handedness of the phase is governed mainly by the pitch of
the Yukawa coil. The
symmetry of the cholesteric phase need not be equivalent
to that of the individual helices but may be different,
depending on the precise value of the coil pitch. Within a certain
interval of the coil pitch, right-handed
Yukawa coils may generate left-handed cholesteric order and vice versa.
 The antagonistic effect of charge-mediated chiral interactions is consistent with experimental observations in {\em M13} virus systems \cite{tombolato-grelet} and
various types of DNA \cite{livolantDNAoverview,tombolatoferrarini}
where left-handed cholesteric phases are formed from right-handed
helical  polyelectrolyte conformations.
Small variations in the shape of the helical coil,
induced by e.g. a change of temperature, may lead to a sense inversion
of the cholesteric helix. Such an inversion has been found in thermotropic (solvent free) polypeptides \cite{watanabe} and cellulose derivatives \cite{yamagishi}, and in mixtures of right-handed cholesterol chloride and left-handed cholesterol myristate \cite{sackmann}.

Our model also predicts that a very small degree of microscopic
helicity is required to maximize the twisting potential of the Yukawa
helices. The optimum is reached when the pitch distance of the coils equals about $ 5 $ times the rod length. 
Mixing stiff helical rods with sufficiently different lengths may lead to a demixing of the
cholesteric phase at high pressures.  The demixing region closes off at a
critical point upon lowering the osmotic pressure. 

The present model could be interpreted as a simple prototype for
complex biomacromolecules such as DNA and {\em fd} which are characterized
by a helical distribution of charged surface groups. Other lyotropic cholesteric systems, such as cellulose and chitin
microfibers in solution could also be conceived as charged rods with
a twisted charge distribution \cite{chitin-revol}.
Small changes in  the internal twist of the fibrils could be induced (e.g. by applying
an external field or varying the temperature) in order to tune the
handedness of the cholesteric phase. This could be of importance
for the use of chiral nanocrystals in optical switching devices and
nanocomposites. We remark that a more accurate description of the
pitch sensitivity,  particularly for DNA systems, could be formulated
by accounting for the steric contributions associated with the
 helical backbone of the chains as well as the influence of chain
 flexibility. This could open up a route towards understanding the unusual
 behaviour of the pitch versus particle and salt  concentration
 as encountered in DNA \cite{livolantDNAoverview,strey-dna} using simple
 coarse-grained models. Furthermore, a quantitative comparison of the
 predicted pitch distances with experimental systems should
 be possible for rigid, slender polyelectrolytes with a well-defined surface charge and
 internal pitch. Most chiral systems studied to date however do not
 fulfill these criteria which makes it difficult to put our predictions to a quantitative test. At
 present, our theory should therefore be considered as a mere qualitative guideline.

Future work will be aimed at studying the pitch sensitivity beyond the
purely Gaussian approximation. This can be done by adopting a numerical
approach to determine the local ODF in a self-consistent way.
This approach could unveil a much more complex relation between the pitch
handedness and system density (or temperature) as suggested by the intricate angle-dependence of the
chiral potential in \fig{twangle}. Such a sense inversion upon changing
temperature has been found theoretically by Kimura \etal  \cite{kimura,kimura2}
within a simple mean-field (Maier-Saupe) treatment of the hard
rod model combined with a Goossens-type chiral potential. It
would be intriguing to see whether a similar effect could be generated
from the present Yukawa model. Investigations along these lines are currently being undertaken.

\acknowledgments

HHW acknowledges the Ramsay Memorial Fellowship Trust for financial support. Funding to the Molecular Systems Engineering group from the Engineering and Physical Sciences Research Council (EPSRC) of the UK (grants GR/N35991 and EP/E016340), the Joint Research Equipment Initiative (JREI)
(GR/M94427), and the Royal Society-Wolfson Foundation refurbishment grant is gratefully acknowledged.

\section*{Appendix: Gaussian averages}

The Gaussian averages required for the calculation of the twist
elastic constant are taken from \olcite{odijkelastic}. We quote them
here:
\begin{widetext}
\bea
\lang \gamma \rang  & \sim & \left ( \frac{\pi}{2} \right )^{1/2}
\left ( \frac{\alpha_{i} + \alpha_{j}}{\alpha_{i} \alpha_{j}} \right)^{1/2}  \\
\lang \gamma^{3} \theta_{i}^{2}  \rang & \sim  & 3 \left (
  \frac{\pi}{2} \right )^{1/2} \left ( \frac{\alpha_{i} +
    \alpha_{j}}{\alpha_{i} \alpha_{j}} \right )^{1/2} \left ( \frac{2
    \alpha_{i} + 5 \alpha_{j}}{\alpha_{i}^{2} \alpha_{j}} \right )  \\
\lang \gamma \theta_{i}^{2} (\theta_{i}^{2} + \theta_{j}^{2} ) \rang 
& \sim & \left ( \frac{\pi}{2} \right )^{1/2} \left ( \frac{6
    \alpha_{i}^{3} + 19 \alpha_{i}^{2}\alpha_{j} + 30 \alpha_{i}
    \alpha_{j}^{2} + 15 \alpha_{j}^{3}}{\alpha_{i}^{5/2}
    \alpha_{j}^{3/2} ( \alpha_{i} + \alpha_{j} )^{3/2} } \right )
\eea
\end{widetext}
\bibliography{rik}
\bibliographystyle{apsrev}

\end{document}